\def\year{2022}\relax
\documentclass[letterpaper]{article}
\usepackage{aaai22} 
\usepackage{times} 
\usepackage{helvet} 
\usepackage{courier} 
\usepackage[hyphens]{url} 
\usepackage{graphicx} 
\usepackage{graphicx} 
\usepackage{natbib} 
\usepackage{caption} 
\usepackage{multirow}
\DeclareCaptionStyle{ruled}%
{labelfont=normalfont,labelsep=colon,strut=off}
\title{Forecasting West Nile Virus with Graph Neural Networks: \\Harnessing Spatial Dependence in Irregularly Sampled Geospatial Data}
\author{
Adam Tonks,\textsuperscript{\rm 1}
Trevor Harris,\textsuperscript{\rm 2}
Bo Li,\textsuperscript{\rm 1}
William Brown,\textsuperscript{\rm 3}
Rebecca Smith \textsuperscript{\rm 3}
}
\affiliations {
\textsuperscript{\rm 1}Department of Statistics, University of Illinois at Urbana-Champaign\\
\textsuperscript{\rm 2}Department of Statistics, Texas A\&M University\\
\textsuperscript{\rm 3}Department of Pathobiology, University of Illinois at Urbana-Champaign\\
aytonks2@illinois.edu,
tharris@stat.tamu.edu,
\{libo, wmbrown, rlsdvm\}@illinois.edu
}

\begin{document}
\maketitle
\begin{abstract}
Machine learning methods have seen increased application to geospatial environmental problems, such as precipitation nowcasting, haze forecasting, and crop yield prediction. However, many of the machine learning methods applied to mosquito population and disease forecasting do not inherently take into account the underlying spatial structure of the given data. In our work, we apply a spatially aware graph neural network model consisting of GraphSAGE layers to forecast the presence of West Nile virus in Illinois, to aid mosquito surveillance and abatement efforts within the state. More generally, we show that graph neural networks applied to irregularly sampled geospatial data can exceed the performance of a range of baseline methods including logistic regression, XGBoost, and fully-connected neural networks.
\end{abstract}

\section{Introduction}

\subsection{West Nile virus}
West Nile virus (WNV) is an arbovirus that causes the disease West Nile fever. The virus was first identified in Uganda in 1937, and subsequently in other central African countries shortly thereafter. The first appearance of WNV in the United States occurred in 1999, and the disease has since become endemic in much of North America. The death rate of those with serious disease is about 10\%. The number of annual deaths across the United States alone has occasionally reached the hundreds, highlighting the urgent need for improved measures to curb WNV transmission. \cite{wnv_info}

\begin{figure}[ht!]
\centering
\includegraphics{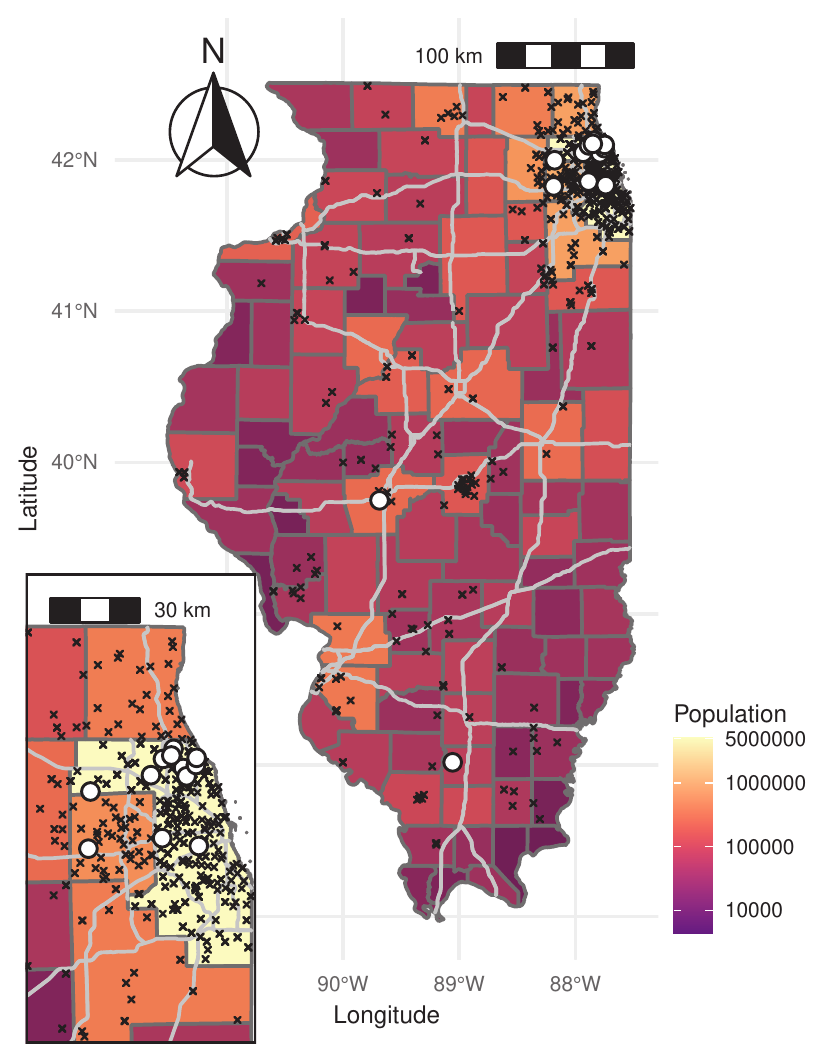}
\caption{Location of traps tested during the week of Monday July 15 2019 to Sunday July 21 2019 in Illinois, with Chicago metropolitan area inset (negative tests indicated as crosses and positive tests as circles)}
\label{il_traps}
\end{figure}

Birds serve as a reservoir for WNV, while certain species of mosquitoes, primarily \textit{Culex}, serve as vectors to transmit WNV from birds to humans. Curbing transmission of WNV via mosquito control remains challenging \cite{zika_detection}, in part due to the difficulty of effective targeting of mosquito eradication techniques such as insecticide, larvacide and bug traps. Such eradication techniques can be highly resource intensive, and judicious selection of deployment location is one way to improve their effectiveness \cite{review_ml_mosquito}. Therefore, accurate short-term forecasting of mosquito populations and WNV disease can aid overall mosquito control efforts. Notably, although the problem of forecasting mosquito populations and disease from trap and weather data has seen extensive application of machine learning methods, no such previous work has emphasized the spatial dependence inherent to the problem \cite{review_ml_mosquito}.

\subsection{Deep learning for geospatial data}
Over the past decade, deep learning methods have been increasingly used for scientific modeling in areas such as particle physics \cite{physics_dl}, drug discovery \cite{drug_dl}, and environmental forecasting \cite{convlstm}. Although a range of machine learning methods, including deep learning methods, have seen applications to forecasting with geospatial data, much of the previous work has either concentrated on object classification in image-like data collected using remote sensing \cite{review_ann_geospatial, keras_spatial}, or employed models that do not inherently account for the spatial dependence of geospatial datasets \cite{review_ml_mosquito}. A notable exception to this is the use of neural networks employing convolution layers, namely convolutional neural networks (CNNs) and graph neural networks (GNNs), to generate nowcasts or forecasts using geospatial data \cite{convlstm, haze_forecasting, traffic_forecasting, gnn_rnn}.

CNNs have been shown to perform well on gridded spatiotemporal data \cite{convlstm, gnn_rnn}, for example popular climate model output dataset products such as CPC Merged Analysis of Precipitation (CMAP). Previous work has shown that CNNs using such data can outperform other statistical models in the nowcasting and forecasting of precipitation and haze, especially when their architectures are further specialized to the problem domain, as is the case with HazeNet \cite{chien_haze} and ConvLSTM \cite{convlstm}. However, CNNs are not applicable to many geospatial problems where spatial processes are irregularly sampled and cannot be directly encoded into a simple lattice. Although it is possible to interpolate observations onto a grid so that existing CNN approaches can be employed, this could lead to propagating errors if the interpolation were inaccurate.

\subsection{Graph neural networks and GraphSAGE}

Irregularly sampled geospatial data can, however, be represented as a spatial graph, which encodes spatial observations as nodes and uses some measure of distance as edge weights. Therefore, GNNs offer an attractive substitute over CNNs, since they can operate directly over arbitrary graphs, such as spatial graphs. GNNs have been previously applied to geospatial data for prediction of place characteristics \cite{gnn_places}, metro station area vibrancy \cite{gnn_vibrancy}, and social media check-ins \cite{gnn_social}, as well as for forecasting of crop yield \cite{gnn_rnn} and traffic \cite{traffic_survey}. Often, we can choose an input spatial graph that clearly reflects the actual locations of observations; this is the case with crop yield data for polygonal regions, where region adjacencies can be translated directly to graph adjacencies, and with traffic data, where the literature already suggests methods for constructing such graphs (most typically using road links for edges) \cite{traffic_survey}.

When there is no clear choice of an input spatial graph, we may always encode observations as a complete graph. However, this may not be the best alternative. We may construct a hierarchy of spatial subgraphs for linking spatial point events, ranging from the complete graph to mutually nearest neighbors \cite{fortin_dale_spatial}. Outside of this hierarchy, the representation of spatial information into graphs that can be used in GNNs remains an area of active research \cite{pe_gnn, spatial_gcn, learning_spatial}. To the best of our knowledge, there is a lack of previous work in exploring the real-world application of GNNs to forecasting problems using irregularly sampled geospatial data encoded as such spatial subgraphs. To this end, we demonstrate an application of GNNs to the forecasting of WNV-positive mosquitoes in Illinois, and show that our model outperforms existing state-of-the-art methods.

GNNs can be seen as generalizations of CNNs, with convolutional operations compatible not only with lattice graphs, but arbitrary ones. Furthermore, these operations are permutationally invariant, meaning that the output of such operations is the same for isomorphic graphs, regardless of the permutation of input nodes \cite{graph_sage}. As a rough example, a properly defined graph convolution would produce the same output regardless of which adjacency matrix is used as input. In contrast, convolutions used in a CNN are not permutationally invariant; reordering image pixels while maintaining the same lattice graph, such as by rotating or mirroring the image, leads to different convolution operation outputs.

GNNs are typically considered within a ``message-passing'' framework, where the neighborhood of each node defines a computational graph from which ``messages'' generated by nearby nodes are received and then ``aggregated'' to compute node embeddings which can be used for prediction \cite{message_passing}. Multiple graph convolution layers can be stacked within a GNN to increase the depth of the computational graph, thus increasing the maximum distance to nodes whose features are considered in generating node embeddings and predictions.

GraphSAGE is a framework proposed in 2017 for efficiently generating node embeddings using node feature information \cite{graph_sage}. In contrast to many other approaches for generating node embeddings, GraphSAGE can operate on previously unseen nodes, including completely unseen graphs. This makes GraphSAGE ideal for our purposes, since input graphs for which we wish to produce predictions correspond to days not included in the training data.

As considered within the message-passing framework, at each layer, the GraphSAGE operator with the mean aggregator applied to node $i$ is defined as
$$\mathbf{h_i'}=\mathbf{W_1}\mathbf{h_i}+\mathbf{W_2}\frac{\sum_{j\in\mathrm{N}(i)}\mathbf{h_j}}{|\mathrm{N}(i)|},$$
where $\mathbf{h_i}$ is the current node embedding, $\mathbf{h_j}$ are the node embeddings of neighboring nodes, and $\mathbf{W_1}$ and $\mathbf{W_2}$ are matrices of trainable weights. In effect, nodes are no longer treated as independent observations using only the $\mathbf{W_1}\mathbf{h_i}$ term as the layer output, as would be the case in a fully-connected neural network. Instead, the GraphSAGE operator combines this term with $\mathbf{W_2}\frac{\sum_{j\in\mathrm{N}(i)}\mathbf{h_j}}{|\mathrm{N}(i)|}$, which is a weight matrix applied to the mean of node embeddings from nodes adjacent to node $i$.

\subsection{Related work}
Previous efforts to forecast mosquito populations and disease can be roughly divided into three different approaches: those using mechanistic models, those using conventional statistical methods, and those using machine learning methods. A recent review examined 221 papers that employed machine learning methods, and summarized a selection of 120 of these. The most popular mosquito genera studied was \textit{Aedes}, appearing in 109 of the papers. \textit{Culex} was studied in 53 of the papers (some of the selected papers studied multiple genera).

Half of these papers utilized geospatial data for their predictions, while the remaining papers analyzed vision, audio, text, and citizen science data. Of these geospatial papers that directly forecasted the spread of mosquito borne diseases, support-vector machines, decision trees and artificial neural networks were utilized with varying degrees of success. The most popular diseases studied were dengue and malaria.

All surveyed work exploring artificial neural networks applied to geospatial data used either fully-connected networks or recurrent neural networks (RNNs). Some of this work utilized synthetic rather than real data \cite{aedes_ai, dl_synthetic}. A study that applied a fully-connected artificial neural network to predict mosquito abundance in urban areas of South Korea varied the learning rate and number of neurons in the hidden layer to find the model with minimum mean square error \cite{korea_ann}.

The use of artificial neural networks to predict malaria dates back to at least 2006 \cite{thailand_ann}. This work predating the era of more sophisticated neural network architectures understandably uses a rather rudimentary fully-connected neural network with just one hidden layer. In later years, neural networks with two hidden layers were explored for predicting malaria in Brazil \cite{brazil_ann}. The same approach was revisited again in more recent years \cite{malaria_analysis}.

Two studies forecasting dengue in Brazil \cite{brazil_lstm} and Kuala Lumpur \cite{kl_lstm} used LSTM models, which are a type of RNN. The Brazil study was notable in using hierarchical clustering as a pre-processing step to determine variables associated with nearby cities to use as features for generating city-by-city forecasts. This is in contrast to other studies that generally did not account for spatial dependence in mosquito abundance or disease. Transfer learning has also been explored in conjunction with LSTM models for dengue forecasting \cite{china_lstm_transfer}.

Although the use of $k$-nearest neighbors in combination with GNNs has been mentioned in the literature \cite{gnn_social}, such an approach has not yet seen widespread real-world applications. In particular, none of the surveyed mosquito abundance and disease prediction papers used machine learning approaches that accounted for the spatial dependence of the data.

\section{Data}

\subsection{Dataset}
The dataset provided by the Illinois Department of Public Health (IDPH) contains $133,867$ observations from 2008 to 2021. Each observation corresponds to the testing of a particular trap on a particular day. These traps are of the gravid type and collect primarily \textit{Culex} mosquitoes, although they do occasionally capture \textit{Aedes albopictus}. Variables include trap latitude and longitude, sample collection day, number of mosquitoes in sample, and test result (positive or negative). Figure \ref{il_traps} shows the trap locations of a particular week in 2019. County populations and Interstate Highways are indicated on the map. It is seen that most trap locations are concentrated around Chicago and other population centers.

\begin{figure}[ht]
\centering
\includegraphics{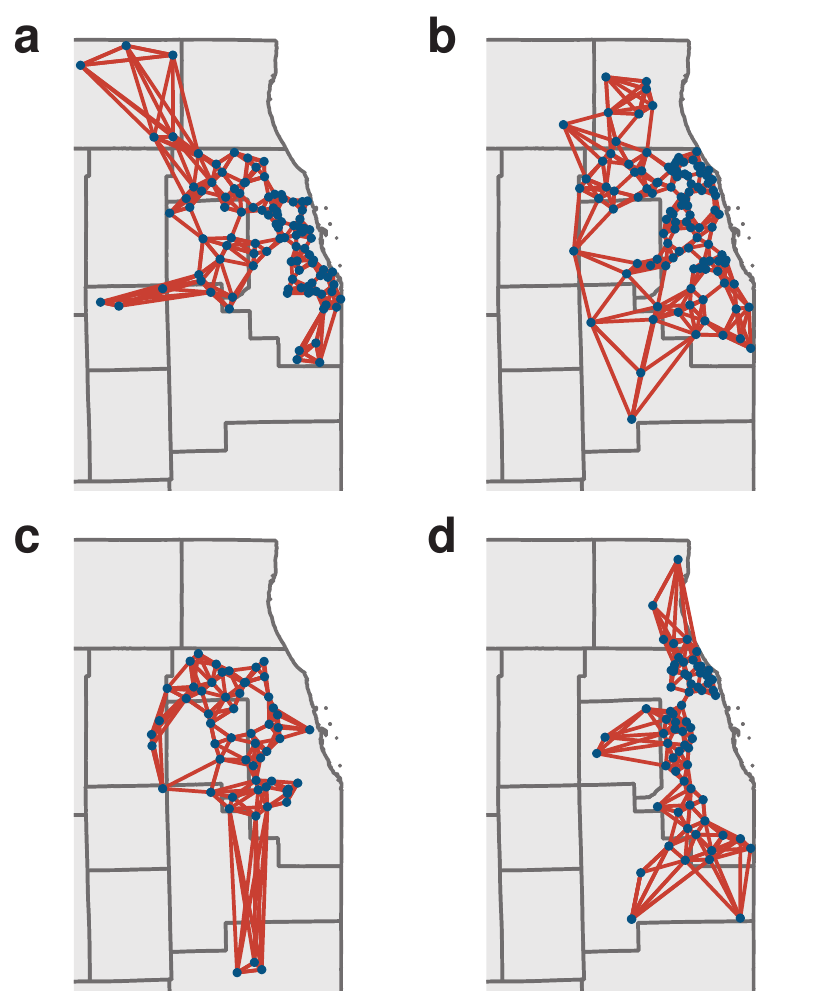}
\caption{GNN input graphs created for (a) Tuesday July 16 2019 (b) July 17 (c) July 18 (d) July 19 using $k$-nearest neighbors with $k=5$, zoomed into the Chicago metropolitan area}
\label{il_graph}
\end{figure}

Since a large number of traps were tested infrequently, sometimes just a few times over the $13$ year span of the dataset, trap locations with a testing frequency less than the $90^{th}$ percentile ($128$ tests) are excluded from our analysis. This corresponds approximately to a minimum of $1$ test each week during the summer mosquito season. This reduces the total number of observations in the dataset from $133,867$ to $81,481$, and unique trap locations from $3332$ to $333$.

\subsection{Formulation of WNV forecasting problem}
Let $P_i(t)$ denote the test result for trap $i$ on day $t$. We let $P_i(t)=1$ if trap $i$ tested positive on day $t$, and $P_i(t)=0$ if trap $i$ tested negative, or was not tested at all, on day $t$.

For any trap $i_0$, we wish to use trap and weather data collected on day $t_0$, or prior, to forecast whether trap $i_0$ will observe a positive reading in the $l^{th}$ week following day $t_0$, for $l$ in the range $1$ to $7$. That is, if we are considering $P_{i_0}(t_0)$, we wish to predict whether any of $P_{i_0}(t_0+7l), P_{i_0}(t_0+[7l+1]), \ldots, P_{i_0}(t_0+[7l+6])$ equal $1$. We consider forecasts up to week $7$. In other words, we wish to predict the variable $A_{i_0}(t_0+7l) = \max(P_{i_0}(t_0+7l), P_{i_0}(t_0+[7l+1]), \ldots, P_{i_0}(t_0+[7l+6]))$ for lead times $l \in \{1, \ldots, 7\}$.

For example, if we are generating forecasts for trap $5$ on day $100$, we wish to make $7$ total forecasts: whether or not a positive test will be observed at that trap within the week spanning days $107$ to $113$, within the week spanning days $114$ to $120$, and so on, up to the week spanning days $149$ to $155$. The features used may include weather and neighboring trap data from day $100$ and any preceding days.

By considering a function of $7$ consecutive days, instead of just a day at a time, we gain a better idea of whether WNV-positive mosquitoes are present in the trap vicinity. This is because for any particular day, a trap only captures a small, biased sample of the mosquito population. Hence considering mosquitoes captured across several days allows for a larger, more representative sample that is still relevant to the time point of interest. Although we could have taken the mean of $P_{i_0}(t_0+7l), P_{i_0}(t_0+[7l+1]), \ldots, P_{i_0}(t_0+[7l+6])$ to achieve smoothing similar to kernel density estimation, taking the maximum instead allows us to interpret $A_{i_0}(t_0+7l)$ simply as the positivity of a mosquito pool aggregated from pools collected on days $t_0+7l$ to $t_0+[7l+6]$.

Our formulation of this forecasting problem is similar to that seen previously in the literature, where a similar smoothing scheme was conducted using a $3$-week moving window to reduce variation in the data resulting from environmental factors that were not of interest, such as moonlight and wind speed \cite{wnv_pred_quebec}.

\section{Methods}

\subsection{Model}
We use a $4$-layer GNN model using GraphSAGE layers with the mean aggregator and no $L^2$ normalization. The model contains a total of $361$ trainable parameters. The first $3$ layers generate $8$-dimensional node embeddings, which are then passed through a ReLU activation function. The last layer generates $1$-dimensional node embeddings that are used as inputs to a sigmoid function to predict the probability of a positive case in the $l^{th}$ week $\hat{P}(A_{i_0}(t_0+7l)=1)$.

\subsection{Graph creation and node features}
A graph is created for each day in the dataset. Each node in the graph represents a trap location, and edges between nodes are determined by $k$-nearest neighbors, thus creating a $k$-nearest neighbor graph ($k$-NNG). Features, such as past trap positivity and weather, are associated with each node. Since the number and location of tested traps differ each day, the graphs also differ by day. The graphs are undirected and unweighted, and may be disconnected.

\begin{figure}[ht]
\centering
\includegraphics{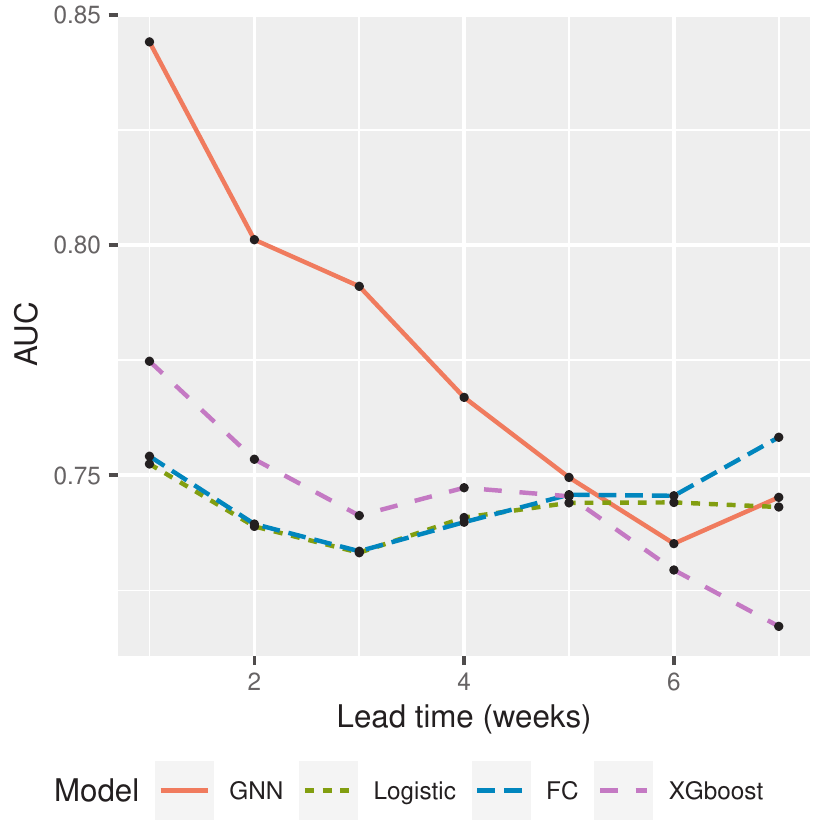}
\caption{AUC values for GNN ($k=5$) and baseline models at increasing lead times}
\label{baseline_comp}
\end{figure}

Edges found using $k$-nearest neighbors that connect two traps at a distance greater than $100$km are excluded from the graphs, to limit the range of the spatial dependence accounted for. We present results using graphs for $k=1, 2, 5, 10, 20$. An example of generated graphs with $k=5$ is depicted in Figure \ref{il_graph}. Distances between trap locations are calculated from their latitudes and longitudes using the Haversine formula.

Recent trap and weather data are associated directly with the nodes as features. For the node corresponding to location $i_0$ and day $t_0$, the feature $A_{i_0}(t_0-6) = \max(P_{i_0}(t_0-6), P_{i_0}(t_0-5), \ldots, P_{i_0}(t_0))$ encodes trap positivity from the past week. The weather features $X_j(t_0)$ encode data for weather variables $j \in \{1, 2, 3\}$, corresponding to cooling degree days, heating degree days and precipitation at Chicago O'Hare International Airport on day $t_0$. These weather variables were found to be statistically significant in related work on WNV and mosquito abundance forecasting \cite{wnv_pred_peel}.

\subsection{Model tuning, training and testing}
Models are trained in PyTorch Geometric on a server running CentOS Linux 7 (Core) utilizing a NVIDIA Quadro GV100 ($32$GB). Training of each GNN model takes approximately $30$s.

The dataset is split into training (years 2008 to 2016), validation (years 2017 to 2018) and test (years 2019 to 2021) sets, covering approximately $8.5$ years, $2$ years and $2.5$ years of data respectively.

The Adam optimizer is used with a learning rate of $0.001$ to minimize cross-entropy loss using a batch size of $100$. Models are trained until no improvement in validation loss is seen in $10$ consecutive epochs, up to a maximum of $1000$ epochs.

Tuning of the model hyperparameters is conducted using the validation set without reference to the test set.

All code and the dataset provided by the IDPH are included in the supplementary material.

\section{Experiments}

\subsection{Baseline models}
The performance of the GNN model is compared to that of several standard baseline machine learning and statistical models. These are logistic regression, a fully-connected neural network, and XGboost. Logistic regression is a commonly used statistical model that models the log-odds of a binary event as a linear combination of a set of features. Furthermore, since we are using the mean aggregator with GraphSAGE, it is equivalent to a $1$-layer GNN model. Fully-connected neural networks are the most basic type of neural network, upon which all other architectures are built upon. We used a $4$-layer network with $8$ neurons per layer and ReLU activation functions, except for the output layer. Finally, XGboost is a popular, state-of-the-art implementation of a decision tree boosting algorithm and is commonly used in machine learning competitions due to its excellent predictive performance \cite{xgboost}.

These models are trained in a similar manner as the GNN model, using early stopping to prevent overfitting. Features are identical, but there is no consideration of the graph structure imposed on the data. The observations are therefore assumed to be independent.

Though this is a spatiotemporal forecasting problem, our focus was on its spatial aspects. Therefore, since the temporal context in our GNN is simply represented by feature values at previous timesteps, we did not include any baseline models that account for the temporal context in a more sophisticated manner, such as specialized autoregressive models.

\subsection{Performance metrics}
The performance metrics primarily considered are accuracy, Brier score, and AUC. Since the models output probability predictions $\hat{P}(A_{i_0}(t_0+7l)=1)$, accuracy is computed by considering the predicted class as that with the larger predicted probability. That is, $\hat{A}_{i_0}(t_0+7l)=1$ if $\hat{P}(A_{i_0}(t_0+7l)=1)>p_t$, and $0$ otherwise, where $p_t=0.5$.

Since accuracy only considers the performance of a model at a particular selected trade-off between sensitivity and specificity, for example that selected here by letting $p_t=0.5$, we primarily consider AUC in assessing model performance. The need to consider performance across a range of trade-offs between sensitivity and specificity is particularly relevant in the context of epidemiology and medicine, and thus has seen widespread usage in these fields. That AUC is superior to accuracy when judged on a number of formal criteria has also been established in the machine learning literature \cite{acc_vs_auc}. For instance, by considering the sensitivity and specificity across a range of threshold probabilities $p_t$, AUC effectively accounts for the ``confidence'' of a class prediction, while accuracy disregards the magnitude of predicted probabilities and considers only on which side of $p_t$ they lie. Finally, AUC has a simple interpretation: given a randomly chosen positive observation and a randomly chosen negative observation, AUC is the probability that the predicted probability of the positive observation being in the positive class is greater than that of the negative observation.

We provide the Brier score, a proper scoring rule, as an additional metric to assess the models. The Brier score is calculated as the mean squared error between predicted probabilities for the positive class and actual class labels. Smaller values indicate better performance.

\subsection{Determining $k$ for input graphs}
\begin{figure}[ht]
\centering
\includegraphics{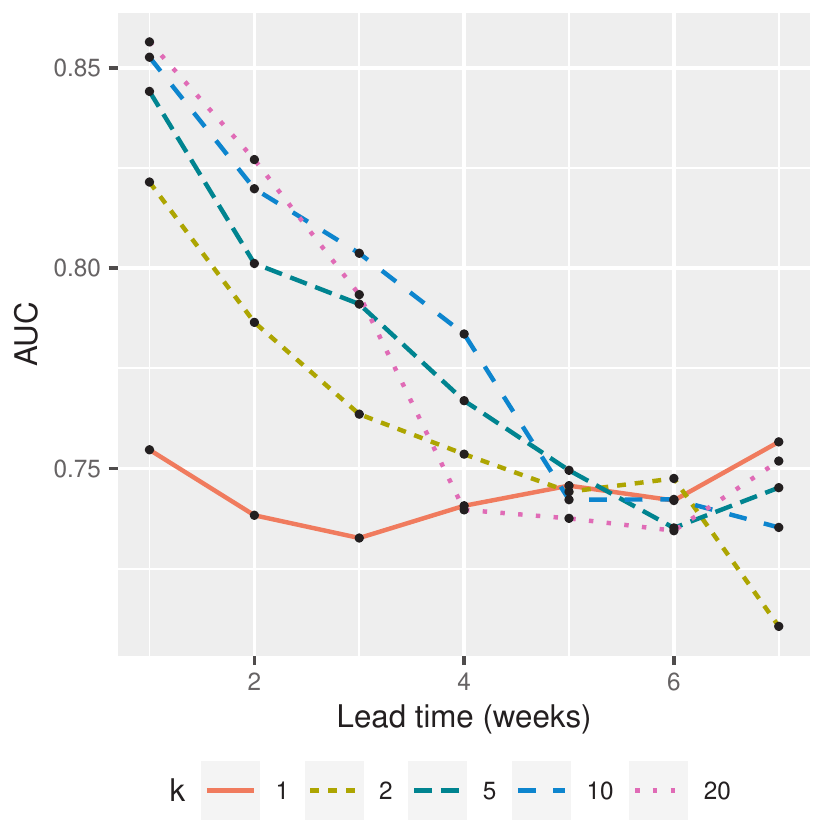}
\caption{AUC values for GNN models using various input graphs at increasing lead times}
\label{graph_k_comp}
\end{figure}

Graphs are created using $k$-nearest neighbors for $k=1, 2, 5, 10, 20$. Graph nodes are associated with both trap and weather data and used as input into the GNN model. The AUC values of using these graphs for various lead times is shown in Figure \ref{graph_k_comp}, and are used to select an appropriate value of $k$ for later model comparisons.

While no single value of $k$ results in model performance greater than other values across all lead times, the AUC values corresponding to $k=5$ and $=10$ are relatively high across all lead times. The sum of AUC across all lead times is also highest for $k=5$ and $k=10$. Therefore, we use GNN models with graphs corresponding to $k=5$ for subsequent comparisons, since doing so offers similar performance but faster computation than $k=10$.

\subsection{Forecasting results}
\begin{figure}[ht]
\centering
\includegraphics[width=0.47\textwidth]{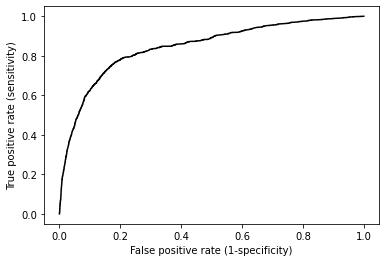}
\caption{ROC curve for GNN ($k=5$) model}
\label{gnn_roc}
\end{figure}

Figure \ref{baseline_comp} and Table \ref{res_tab_main} indicate that at lead times of $1$ to $5$ weeks, the GNN model outperforms all other baseline models in terms of AUC and Brier score. The associated ROC curve is illustrated in Figure \ref{gnn_roc}. The performance difference is most substantial at the smallest lead times and decreases with increasing lead time. At lead times of $6$ and $7$ weeks, the GNN model outperforms XGboost but is outperformed by logistic regression and the fully-connected neural network model, though all models perform very similarly with the exception of XGboost.

\begin{table}[ht]
    \centering
    \begin{tabular}{p{0.7cm}|p{0.95cm}|p{1cm}|p{1cm}|p{1cm}|p{1cm}}
        Lead time & Model type & Acc-uracy & Brier score & AUC & AUC SD \\ \hline
        \multirow{4}{*}{1} & Log. & 0.8686 & 0.1058 & 0.7525 & 0.0001 \\
        & FC & 0.8720 & 0.1062 & 0.7542 & 0.0012 \\
        & XGb & 0.8710 & 0.1056 & 0.7748 & 0.0000 \\
        & GNN & \textbf{0.8776} & \textbf{0.0926} & \textbf{0.8441} & 0.0024 \\ \hline
        \multirow{4}{*}{2} & Log. & 0.8600 & 0.1124 & 0.7389 & 0.0000 \\
        & FC & 0.8596 & 0.1127 & 0.7394 & 0.0007 \\
        & XGb & 0.8605 & 0.1155 & 0.7535 & 0.0000 \\
        & GNN & \textbf{0.8634} & \textbf{0.1058} & \textbf{0.8012} & 0.0030 \\ \hline
        \multirow{4}{*}{3} & Log. & 0.8439 & 0.1171 & 0.7332 & 0.0001 \\
        & FC & 0.8464 & 0.1178 & 0.7335 & 0.0070 \\
        & XGb & 0.8491 & 0.1210 & 0.7413 & 0.0000 \\
        & GNN & \textbf{0.8494} & \textbf{0.1117} & \textbf{0.7910} & 0.0033 \\ \hline
        \multirow{4}{*}{4} & Log. & 0.8464 & 0.1206 & 0.7408 & 0.0001 \\
        & FC & 0.8444 & 0.1210 & 0.7398 & 0.0024 \\
        & XGb & \textbf{0.8497} & 0.1213 & 0.7473 & 0.0000 \\
        & GNN & 0.8457 & \textbf{0.1178} & \textbf{0.7669} & 0.0066 \\ \hline
        \multirow{4}{*}{5} & Log. & \textbf{0.8464} & 0.1206 & 0.7440 & 0.0002 \\
        & FC & \textbf{0.8464} & 0.1203 & 0.7458 & 0.0060 \\
        & XGb & 0.8344 & 0.1246 & 0.7455 & 0.0000 \\
        & GNN & \textbf{0.8464} & \textbf{0.1199} & \textbf{0.7496} & 0.0138 \\ \hline
        \multirow{4}{*}{6} & Log. & \textbf{0.8464} & 0.1196 & 0.7441 & 0.0559 \\
        & FC & \textbf{0.8464} & \textbf{0.1194} & \textbf{0.7456} & 0.0398 \\
        & XGb & \textbf{0.8464} & 0.1258 & 0.7295 & 0.0000 \\
        & GNN & \textbf{0.8464} & 0.1201 & 0.7352 & 0.0067 \\ \hline
        \multirow{4}{*}{7} & Log. & \textbf{0.8464} & 0.1195 & 0.7431 & 0.0370 \\
        & FC & \textbf{0.8464} & \textbf{0.1180} & \textbf{0.7583} & 0.0445 \\
        & XGb & 0.8362 & 0.1278 & 0.7172 & 0.0000 \\
        & GNN & \textbf{0.8464} & 0.1191 & 0.7452 & 0.0474 \\
    \end{tabular}
\caption{Metrics for logistic regression, fully-connected neural network, XGboost and GNN ($k=5$) models at various lead times (in weeks)}
\label{res_tab_main}
\end{table}

At a lead time of $1$ week, AUC is $8.944\%$ greater for the GNN model than XGboost, the next-best model. At a lead time of $2$ weeks, AUC is $6.330\%$ greater. At the largest lead times of $6$ to $7$ weeks, the fully-connected neural network appears to be the best model in terms of AUC and Brier score, but with an AUC value greater than the GNN model by only $1.415\%$ and $1.758\%$ for lead times of $6$ and $7$ weeks, respectively.

Trap data is more informative of the class labels at the smallest lead times, while weather data is more informative at the greatest lead times. Therefore, we hypothesize that at the smallest lead times, the GNN model is able to capitalize on the spatial dependence of the trap data, unlike the baseline models. However, at the largest lead times, such an advantage no longer exists, and hence all models have roughly similar forecasting performance while relying on the weather data. We investigate this hypothesis further in the following subsection.

We also note that although model accuracies for lead times greater than approximately $4$ weeks do not deviate substantially from those found using a ZeroR model, where all predictions are simply the dataset's majority class, all AUC values are significantly greater than $0.5$. This indicates that these models perform better than one that simply predicts ``no WNV'' for all observations, at least at particular trade-offs between sensitivity and specificity. In addition, the class imbalance further renders AUC as a better performance measure than accuracy in this scenario. In more formal terms, class imbalance causes AUC to be more ``discriminating'' than accuracy, in the sense that AUC can often correctly distinguish between the performance of two models when accuracy cannot \cite{acc_vs_auc}.

\subsection{Use of trap versus weather data}
\begin{figure}[ht]
\centering
\includegraphics{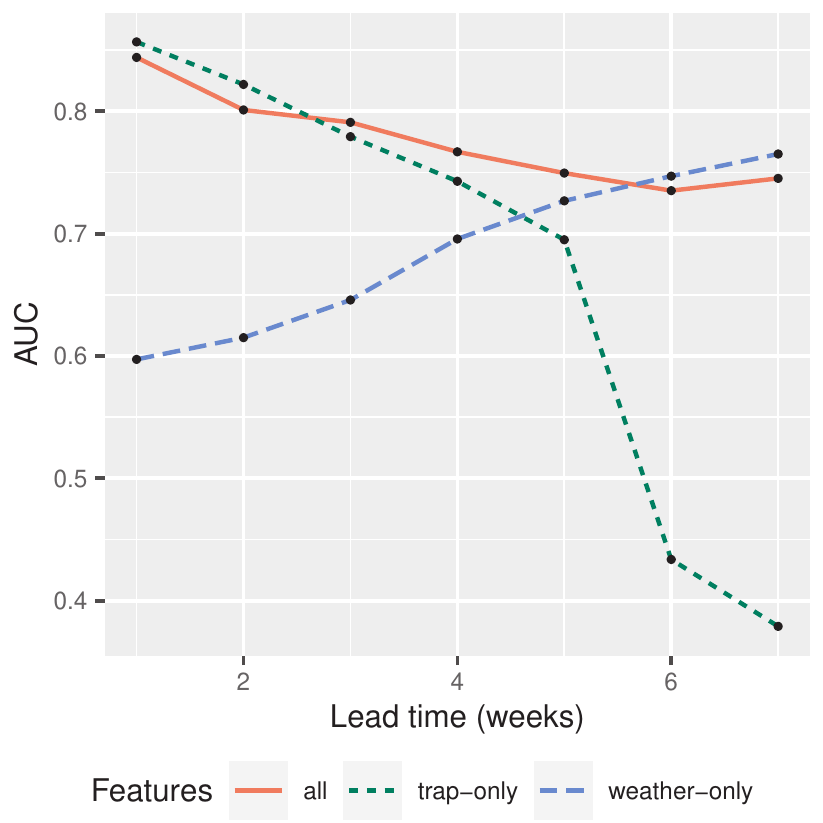}
\caption{AUC values for GNN ($k=5$) model using various features at increasing lead times}
\label{gnn_features_comp}
\end{figure}

Figure \ref{gnn_features_comp} shows the AUC values of the GNN model using trap and weather data, trap data only, and weather data only at increasing lead times. At the smallest lead times, the AUC values are substantially greater using trap data only rather than using weather data only. At a lead time of $1$ week, the AUC value is $0.8568$ using trap data only and $0.5972$ using weather data only, which is a $30.30\%$ decrease.

On the other hand, the performance difference is greater at the largest lead times, but the relationship is reversed. At a lead time of $7$ weeks, the AUC value is $0.7651$ using weather data only and $0.3790$ using trap data only, which is a $50.46\%$ decrease. Notably, the performance of the model using trap data only in this case is below that of a no-skill model with AUC equal to $0.5$. This implies that the trap-only model could simply be equivalent to a no-skill model on the time period covered by the test set, but variance in the test set leads to an AUC value somewhat close, but not equal to $0.5$.

It is not wholly surprising that the trap data $A_{i_0}(t_0-6)$ is best at forecasting $A_{i_0}(t_0+7l)$ for small lead times $l$, because it is expected that two observations tend to be more correlated when they are close in time and space. That weather data plays a stronger role in prediction with larger lead time could be a signal of a cumulative and delayed effect of weather on mosquito populations and disease. The average \textit{Culex} mosquito life cycle is approximately $6$ weeks, so the effects of current weather on mosquito reproduction and development manifest in adult populations approximately $6$ weeks into the future \cite{culex_lifecycle}.

Somewhat surprisingly, at both the smallest and largest lead times, it is not the model using both trap and weather data that performs the best, but rather the trap-only and weather-only models, respectively. At the smallest lead times, we speculate that the variance in the labels explained by the weather data is already almost fully explained by the trap data, so adding weather data to the trap-only model only adversely increases model complexity and hence overfitting on the training set. A similar explanation applies to the largest lead times.

At the intermediate lead times of $3$ to $5$ weeks, the AUC for the model using both trap and weather data is marginally greater than that for either model using trap or weather data alone.

\section{Conclusion}

Previous approaches to forecasting mosquito populations and disease do not effectively account for the spatial dependence of geospatial data. This motivates our employment of a GNN model to forecast WNV in Illinois by constructing input graphs using $k$-nearest neighbors. Our model outperforms several baseline models across a number of metrics for short-term forecasting at lead times of $1$ to $5$ weeks, demonstrating that GNNs are a powerful and practicable approach for WNV forecasting. Future work could include an implementation of our methods in cooperation with public health departments using real-time data to further realize the potential social impact of improved WNV forecasting in mosquito abatement efforts. Additionally, application of a GNN model to mosquito disease forecasting in other regions, or to other vector borne diseases, could be explored. Methods of graph creation beyond $k$-nearest neighbors and the marginal benefit of incorporating RNN units into the model in such contexts may also be of interest.

\section{Acknowledgments}

This material is based upon work supported by the National Science Foundation under Grant No. NSF-DMS-1830312. Any opinions, findings, and conclusions or recommendations expressed in this material are those of the authors and do not necessarily reflect the views of the National Science Foundation.

This publication was supported by Cooperative Agreement \#U01 CK000505, funded by the Centers for Disease Control and Prevention. Its contents are solely the responsibility of the authors and do not necessarily represent the official views of the Centers of Disease Control and Prevention or the Department of Health and Human Services.

\bibliography{aaai22}
\end{document}